\documentclass[3p,times,twocolumn]{elsarticle}

\usepackage{ecrc}

\volume{00}

\firstpage{1}

\journalname{Nuclear Physics B Proceedings Supplement}

\runauth{E. Parizot}

\jid{nuphbp}

\jnltitlelogo{Nuclear Physics B Proceedings Supplement}

\usepackage{amssymb}
\usepackage[figuresright]{rotating}

\begin{document}

\begin{frontmatter}

\dochead{}

\title{Cosmic Ray Origin: Lessons from Ultra-High-Energy Cosmic Rays and the Galactic/Extragalactic Transition}

\author{Etienne Parizot}

\address{Laboratoire Astroparticule et Cosmologie, Universit\'e Paris Diderot/CNRS, 10 rue A. Domon et L. Duquet, 75205 Paris Cedex 13, France}

\begin{abstract}
We examine the question of the origin of the Galactic cosmic-rays (GCRs) in the light of the data available at the highest energy end of the spectrum. We argue that the data of the Pierre Auger Observatory and of the KASCADE-Grande experiment suggest that the transition between the Galactic and the extragalactic components takes place at the energy of the ankle in the all-particle cosmic-ray spectrum, and at an energy of the order of $10^{17}$~eV for protons. Such a high energy for Galactic protons appears difficult to reconcile with the general view that GCRs are accelerated by the standard diffusive shock acceleration process at the forward shock of individual supernova remnants (SNRs). We also review various difficulties of the standard SNR-GCR connection, related to the evolution of the light element abundances and to significant isotopic anomalies. We point out that most of the power injected by the supernov\ae~in the Galaxy is actually released inside superbubbles, which may thus play an important role in the origin of cosmic-rays, and could solve some persistent problems of the standard SNR-GCR scenario in a rather natural way.
\end{abstract}


\end{frontmatter}


\section{Introduction}
\label{sec:intro}

The main reason why the origin of cosmic rays (CRs) is still unknown, one century after their discovery, is that they are charged nuclei isotropized by the turbulent magnetic field in the Galaxy to such a high degree that their observed flux is essentially identical in all directions, with no sources or decisive hot spots identified in any region the sky (cf. below, Sect.~\ref{sec:UHECRs}).

This is all the more frustrating that CRs play a central role in what may be called the Galactic ecology. They are indeed one of the main components of the Galaxy, with an energy density comparable to that of the magnetic field and of stellar light, which makes them key actors of the dynamical equilibrium between the different phases of the interstellar medium (ISM). Galactic cosmic rays (GCRs) play a major role in the heating and ionization of the ISM, and therefore also in the regulation of star formation. In addition, they take part in the complex chemical processes at work in the ISM, and they are responsible for the nucleosynthesis of the so-called light elements (lithium, beryllium and boron). Most $^{7}$Li nuclei, and every nucleus of $^{6}$Li, Be and B in the Galaxy are thought to have been produced by the interaction of CRs with the ISM over the lifetime of the Galaxy.

The longstanding quest for the origin of CRs is not only motivated by the desire to obtain a more complete description of the Galactic ecology. It is also related to the wish to better understand particle acceleration in general, which is a key ingredient of the modeling of high-energy astrophysical sources. These sources are generally studied through their multi-wavelength, non-thermal emission, resulting from the interaction of energetic particles accelerated \emph{in situ} with the ambient matter, photons or magnetic field.

In most cases, the locally accelerated particles exhibit a power-law spectrum which is reminiscent of that observed for the GCRs (after deconvolution of the CR propagation effects). However, one should keep in mind that being able to accelerate nuclei with a power-law spectrum is not enough to make a source or a class of sources an important contributor to the GCRs. Solving the mystery of the CR origin will require understanding the CR phenomenon globally, in all its aspects, which includes the energy spectrum and maximum energy, as well as the composition and anisotropies over the whole range of energies, and in addition their contribution to the above-mentioned processes. In turn, the study of these processes influenced by the GCRs can provide important information and constraints about their sources, which may not have received enough attention yet.

Another way to obtain additional constraints about the cosmic-ray sources is to study the highest-energy end of the spectrum, where the acceleration mechanisms appear to be the most challenging. While the so-called ultra-high-energy cosmic rays (UHECRs), with energies above $\sim 10^{19}$~eV, are most likely to have an extragalactic origin, their much larger magnetic rigidity than the low-energy GCRs may allow the observation of distinct sources in the sky, and provide more direct information about their nature. Moreover, the study of the transition between the Galactic and extragalactic cosmic rays (EGCRs) is important to derive constraints about the highest-energy particles accelerated in our Galaxy, which might in turn challenge the most popular scenarios for the origin of GCRs.

In this paper, we briefly review some of the implications of the currently available data about high-energy cosmic rays, and discuss their relevance to the general problem of identifying the origin of GCRs. We also call into play the fact that the main part of the supernova power injected in the Galaxy is released inside superbubbles, where collective acceleration processes may modify the standard view regarding the link between supernova remnants (SNRs) and the GCRs, which we refer to as the SNR-GCR connection.

\section{UHECRs: basic phenomenology and key results}
\label{sec:UHECRs}

The central element in the phenomenology of UHECRs is the so-called GZK effect, described by Greisen \cite{Greisen1966} and by Zatsepin and Kuzmin \cite{ZatsepinKuzmin1966} just a few weeks after the discovery of the cosmological microwave background (CMB). The GZK effect follows from the interaction of the UHECRs with the extragalactic photon backgrounds. In the case of UHE protons, the dominant interaction is with the CMB photons: it leads to the production of electron-positron pairs for protons with energies above $\sim 10^{18}$~eV, and to the production of pions for protons above $\sim 5\times 10^{19}$~eV. These photo-production processes are of course accompanied with energy losses. In the case of UHE nuclei, the dominant interaction is with the CMB as well as with the infrared background (depending on the energy range). The resulting photo-dissociation leaves the incoming UHE nuclei with one or a few nucleons less, at essentially the same Lorentz factor. While propagating across the universe, the UHE nuclei are thus subject to a photo-erosion which reduces their mass number as well as their energy. For a detailed review on UHECR propagation, see e.g. \cite{Allard:2012} and references therein.

The basic consequence of these processes is the existence of an energy-dependent and mass-dependent cosmic horizon, beyond which the particles of a given energy cannot propagate without losing a significant fraction of their initial energy, as well as being photo-dissociated into low mass nuclei (and eventually into protons), in the case of heavier nuclei. Conversely, an observer on Earth (or at any place in the universe) can only see UHECR sources located within a horizon whose radius depends on the nuclear type and energy of the injected particle. This dependence determines what may be called \emph{the horizon structure} of the UHECRs. It can be easily computed through Monte-Carlo propagation codes, provided that the interaction cross sections and the photon backgrounds are known (with some dependence also on the assumed source distribution around the observer)~\cite{Harari+2006,Globus+2008,Kachelriess+2009,DeDomenicoInsolia2013}.

\begin{figure*}[ht]
\begin{center}
~\hfill\includegraphics[width=0.4\linewidth]{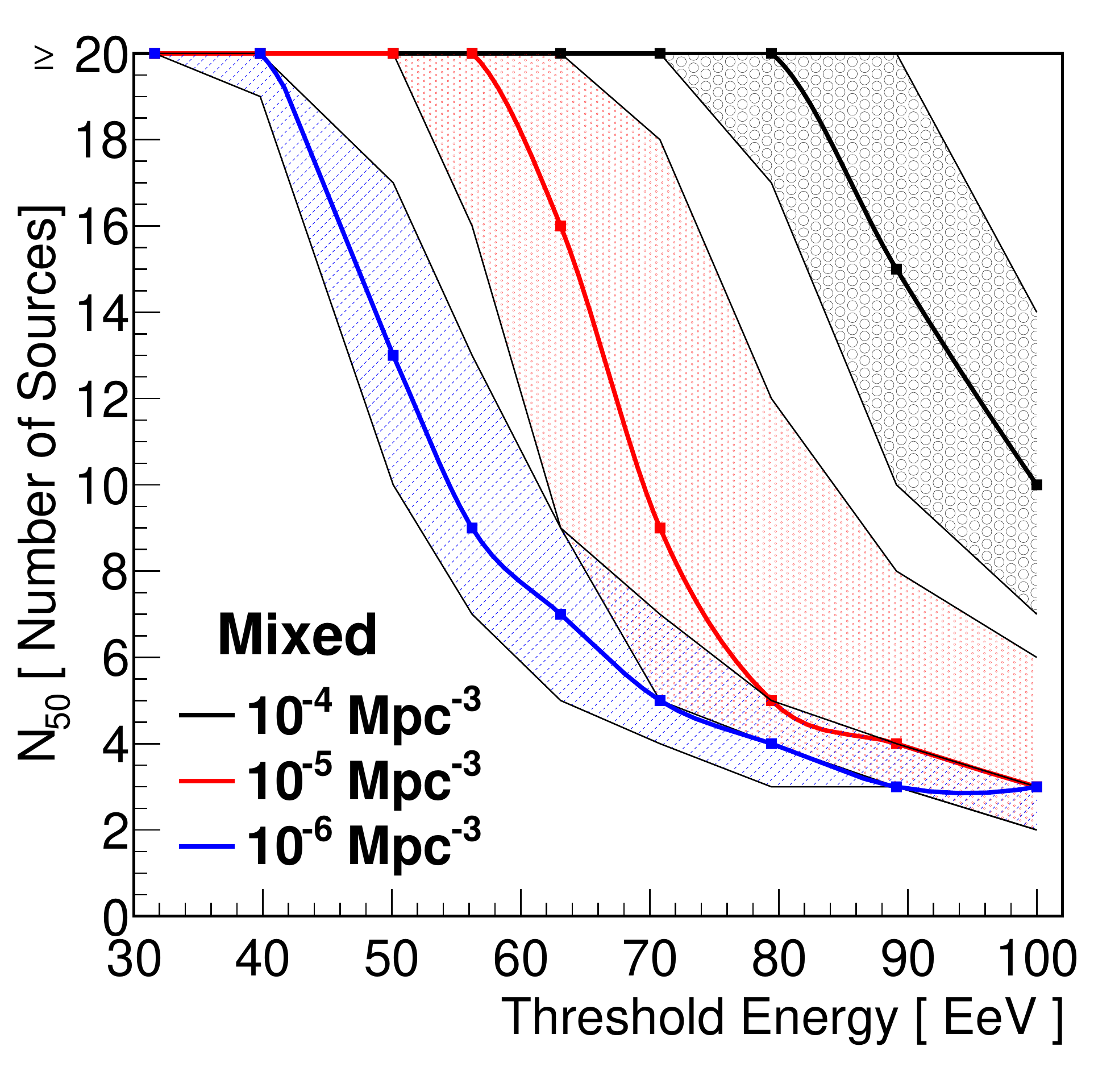}\hfill
\includegraphics[width=0.4\linewidth]{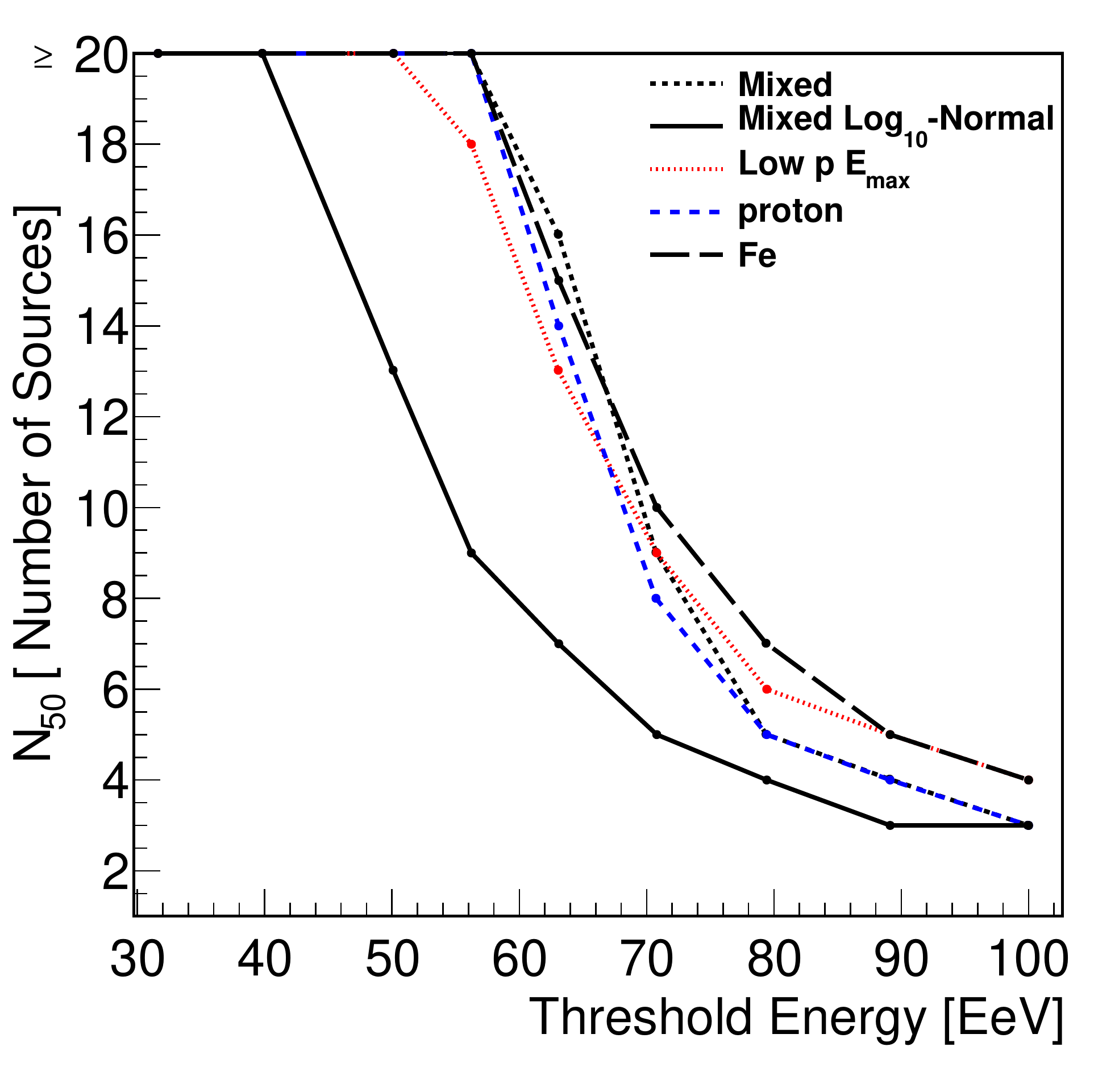}\hfill~
\caption{Number of sources contributing to the flux of UHECRs observed on Earth above a given energy $E$, as a function of that energy (in abscissa), for different UHECR source scenarios. Left: for a mixed-composition scenario, assuming different source densities; right: for different source compositions, with a source density of $10^{-5}\,\mathrm{Mpc}^{-3}$ (see \cite{Blaksley+2013} for details).}
\label{fig:sourceNumber}
\end{center}
\end{figure*}

Because of the GZK effect, the number of sources contributing to the observed UHECR flux decreases with increasing energy, as illustrated in Fig.~\ref{fig:sourceNumber} for different UHECR source models. As can be seen, this decrease is very sharp around 50--80~EeV, which corresponds to the onset of the photo-pion production in proton-CMB interactions, or to the photo-dissociation of Fe nuclei, if the UHECRs are dominated by heavier nuclei (see \cite{Blaksley+2013} for more details on UHECR source statistics). This sharp reduction of the number of contributing sources results in a sharp decrease of the observed UHECR flux, referred to as the GZK cutoff, above $\sim 6\times10^{19}$~eV (see \cite{KoteraOlinto2011} and references therein for a recent review on UHECRs).

Such a cutoff is indeed observed in the UHECR data \cite{HiRes:2008a,Auger:2008a,Auger:2010a,TA:2013a}. However, since the horizon structure happens to be very similar for protons and for Fe nuclei, the high-energy end of the cosmic-ray spectrum can be fitted with a wide range of astrophysical models, with a dominant component of protons or with heavier nuclei, so that it cannot be used to gain any decisive information about the sources. In addition, the observed cutoff in the spectrum may also be the mere signature of the maximum energy reached by the accelerated particles in the sources.

More information has been provided by the composition measurements and the study of UHECR anisotropies. The Auger data \cite{Auger:2010c} have revealed a change in the properties of the UHECR-induced atmospheric showers, which is most naturally interpreted as a change in the composition of the cosmic rays, from a light composition (typically dominated by protons) to a heavy composition, in an energy range between a few $10^{18}$~eV and a few $10^{19}$~eV.

Such a behavior could be a natural consequence of the existence of a relatively low energy cutoff in most UHECR sources. This is the central idea of the so-called ``low proton $E_{\max}$'' models, in which the sources accelerate protons up to a maximum energy $E_{\max}(\mathrm{p})$ of the order of 4--10~EeV, while the heavier nuclei with charge number $Z_{i}$ exhibit a cut-off at the same rigidity, i.e. at an energy $Z_{i}$ times larger \cite{Allard+2008, Aloisio+2011, Allard:2012}. In such scenarios, the cutoff observed in the UHECR spectrum above 60--80~EeV would be essentially the GZK cutoff of the Fe nuclei, since the intermediate nuclei between protons and Fe have a GZK cutoff at lower energy.

The other important piece of information provided by the Auger data and the Telescope Array (TA) data is the absence of any clear anisotropy, notably on small angular scales, which would give direct information about the UHECR sources, or at least allow one to isolate the main sources over the sky. Even though the total exposure of the sky in the Northern hemisphere by TA is smaller than that in the Southern hemisphere by Auger, the most significant structure in the UHECR arrival direction distribution seems to be the so-called TA hot-spot above 57~EeV, with 5.1$\sigma$ pretrial significance, and 3.4$\sigma$ significance after penalizing for the scanning in position \cite{TAHotSpot:2014}. However, the corresponding possible enhancement of the flux in a given direction has an angular scale of the order of 20 degrees, which currently does not allow to identify a source, nor to attribute this putative excess, if confirmed, to a single source or localized cluster of sources. The Pierre Auger Collaboration (hereafter ``Auger'') has also reported an excess of correlation between the arrival direction of UHECRs above $\sim 55$~EeV and the position of nearby AGNs, on a small angular scale of 3 degrees \cite{Auger:2007a,Auger:2010b}. However, the corresponding signal could not reach a high significance level, and did not increase in the following years, despite the large increase in statistics. Moreover, no excess in the auto-correlation function of the UHECR arrival directions could be detected on small-angular scales, where one would expect to detect multiplets of protons coming from single sources. This gives additional support to the interpretation of the composition data in terms of a transition towards heavy nuclei at the highest energies, especially if one keeps in mind that the most intense sources of UHECRs must already have contributed several events above 60~EeV \cite{Blaksley+2013}: these would show as actual small angular scale clusters in the Auger and/or TA sky maps if most of them were protons.

\section{About the GCR/EGCR transition}
\label{sec:transition}

\begin{figure*}[ht]
\begin{center}
\includegraphics[width=0.24\linewidth]{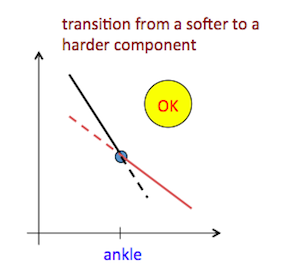}\hfill\includegraphics[width=0.24\linewidth]{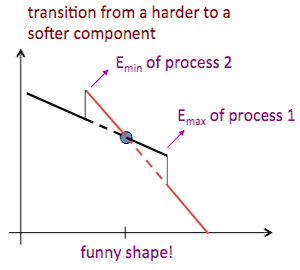}
\includegraphics[width=0.24\linewidth]{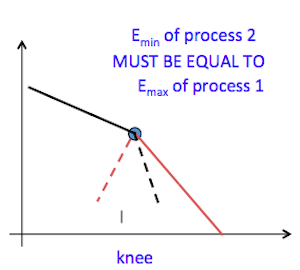}\hfill\includegraphics[width=0.24\linewidth]{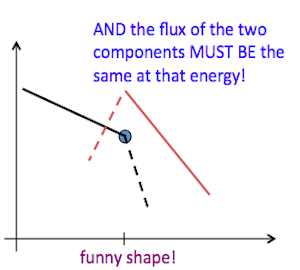}
\caption{Illustration of different types of transitions between two components. If the first one is softer than the second, an ankle is produced. If the second one is harder than the first, no smooth transition is obtained, unless the second one only starts above a given energy, which must happen to be where the first one ends, and at this coincidental energy, the two components must in addition have a similar flux.}
\label{fig:kneeTransition}
\end{center}
\end{figure*}

The above interpretation of the UHECR data in terms of a low-energy cutoff of the protons in the sources should be regarded more as a framework than as a fully definite model. First, it should not be expected that all sources behave in exactly the same way, with the same spectrum, maximum energy and composition, as usually assumed, lacking better knowledge. Second, the source distribution in the vicinity of the Earth -- which includes not only the local source density, but also the actual location and intrinsic power of the individual sources --, plays a role in shaping the UHECR spectrum and influencing the composition and the distribution events over the sky (see \cite{RouilleDOrfeuil+2014} for further discussion on the so-called cosmic variance). However, it is likely that the low proton $E_{\max}$ framework catches some important aspects of the UHECR phenomenology, as revealed by the most recent data.

One important teaching of this framework is that the source spectrum of the extragalactic UHECRs is probably much harder than what had been derived under the oversimplifying assumption of a pure-proton model. Assuming that the source spectrum can be approximated by a power-law, the logarithmic index of this power law should be smaller than 2, and possibly as low as 1.5 \cite{Allard:2012, RouilleDOrfeuil+2014, Aloisio+2013}.

An important consequence of such a hard spectrum is that the extragalactic component responsible for the bulk of the UHECRs above, say, $10^{19}$~eV, cannot dominate the cosmic ray flux much below the spectral break referred to as the ankle, at $\sim 3\times 10^{18}$~eV (see also \cite{Allard+2005,Allard+2007}). This has a direct implication for the sources of the GCRs: at least some of the cosmic-ray sources in our Galaxy must be able to accelerate particles up to $\sim 3\times 10^{18}$~eV or above. This is very challenging for most of the sources investigated in the case of protons. However, if the highest-energy end of the GCR spectrum is dominated by heavy nuclei, as suggested by the available data, the maximum energy reached by the protons in the GCR sources does not need to be so large. In the most standard acceleration scenarios, the maximum energy of different nuclei corresponds to the same maximum \emph{rigidity}, and is thus proportional to the charge, $Z$, for fully ionized nuclei. Among the heavy nuclei at the high-energy end of the GCR spectrum, the most abundant are expected to be Fe nuclei. The existence of Galactic Fe nuclei at the ankle, i.e. at $\sim 3\times 10^{18}$~eV, would thus imply that (at least some) GCR sources accelerate protons up to $\sim 10^{17}$~eV (or possibly above).

The question of the GCR/EGCR transition thus appears as an important question for the understanding of both the GCRs and the UHECRs.

In this respect, an interesting piece of information is added by the observational results recently published by the KASCADE-Grande collaboration (\cite{KG:2011,KG:2013}, and see also \cite{Bertaina:2014}). In their data, they are able to select atmospheric showers with characteristics (muon number vs. size) which make it more probable that they have been induced by a light, rather than a heavy nucleus. According to the latest KASCADE-Grande data, when selecting the CR showers which are induced by the lightest particles (mostly protons, with possibly some fraction of He nuclei), an ankle-like feature is observed at $\sim 10\,^{17}$~eV. This is exactly where one would expect a transition between the Galactic and the extragalactic protons, if the overall GCR/EGCR transition were located at the ankle, with a dominant Galactic Fe component at this energy.

Now, let us try and summarize the situation relating to the GCR/EGCR transition.

In its most usual representation, the global cosmic ray spectrum is described as a power law, $E^{-x}$, with a logarithmic index $x \sim 2.7$ below the knee, then $x \sim 3.0$ above the knee (and apparently an additional ``knee-like'' steepening around $10^{17}$~eV, with $x\sim 3.3$), and finally a hardening at the ankle, with a more complex shape up to the GZK cutoff.

When considering the GCR/EGCR transition, it is natural to expect a feature to be seen in the spectrum in the energy range where it occurs. Moreover, it is natural to expect this feature to be an ankle-like feature, rather than a knee-like feature. This is the reason why the ankle, around $3\times 10^{18}$~eV, has been proposed since its discovery as the signature of the GCR/EGCR transition. In this interpretation, the steeply falling spectrum dominated by the Galactic particles is simply overtaken by the harder spectrum (e.g. because of the absence of an energy-dependent confinement in the Galaxy) dominated by the extragalactic ones. No particular fine-tuning or adjustment is required in this case: the flux ratio between the two components evolves steadily, and the transition simply occurs at the energy where the second component happens to exceed the first one.

On the contrary, it is \emph{a priori} unlikely that the transition between two components occurs at a knee-like feature, where the second component is steeper than the first one. This would indeed require that the first component actually cuts off at the energy of the transition (otherwise its harder spectrum would continue to dominate), and that the second component actually begins at this energy (otherwise its flux would dominate already at lower energy, given its steeper spectrum). Moreover, the flux of the two components is required to be comparable at that very energy where the first one happens to end, and the second one to begin. Otherwise, there would be either a sudden drop or a sudden bump in the CR spectrum (see Fig.~\ref{fig:kneeTransition}). This seems very unlikely, if the two components are not physically related, like in this case where the first one is Galactic, and the second one extragalactic.

Now, let us consider the situation of the cosmic-ray spectrum:
\begin{enumerate}
\item we know that the low-energy cosmic rays have a Galactic origin;
\item we know that the highest-energy cosmic rays (almost certainly) have an extragalactic origin.
\end{enumerate}
Therefore, a transition must occur somewhere, and this is likely to appear in the spectrum -- \emph{a priori} -- as an ankle-like feature. Now, as a matter of fact:
\begin{enumerate}
\setcounter{enumi}{2}
\item an ankle-like feature is seen in the spectrum, around $3$~EeV.
\end{enumerate}

\begin{figure*}[ht]
\begin{center}
\includegraphics[width=0.8\linewidth]{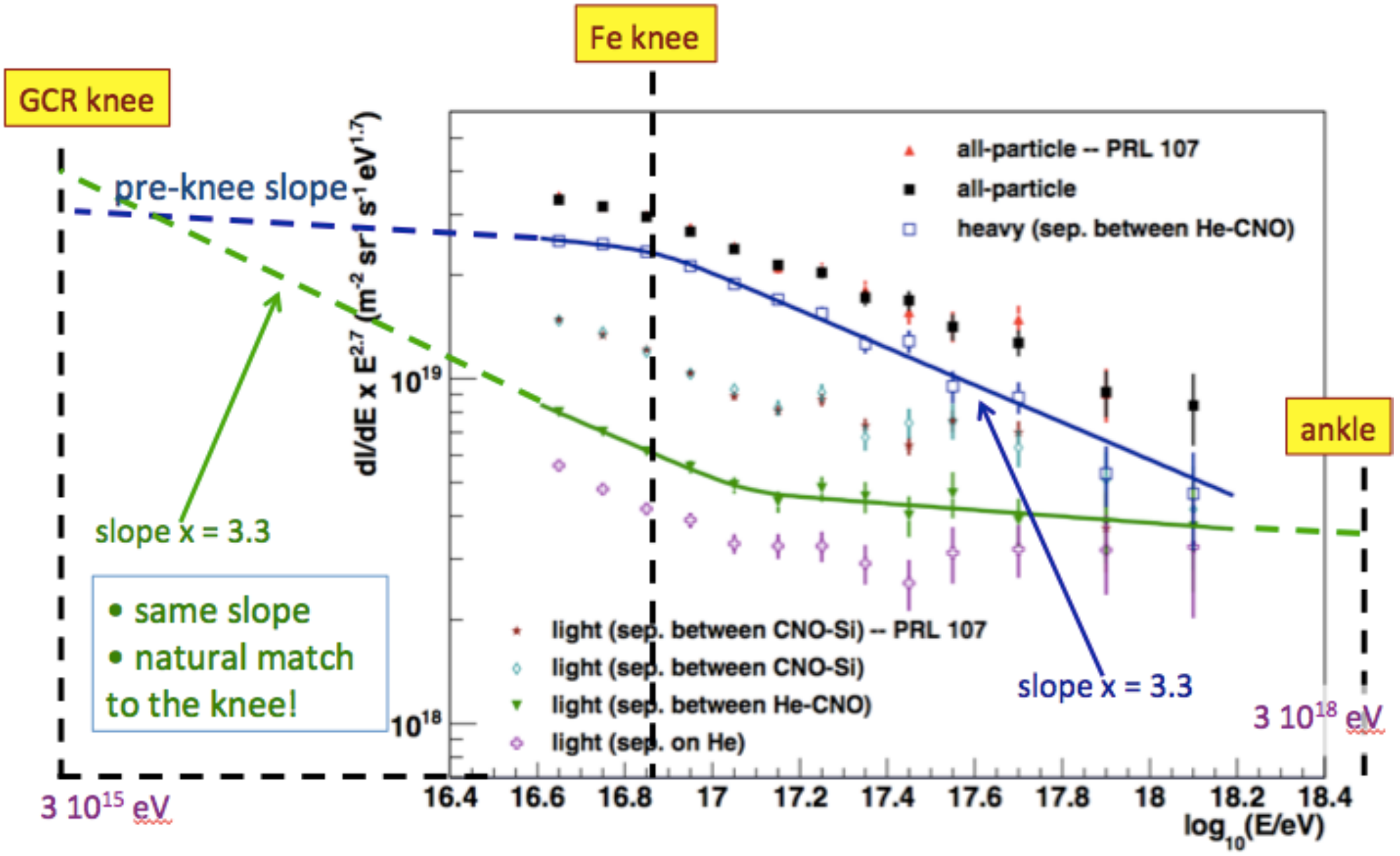}
\caption{Global picture of the possible transition from Galactic to extragalactic cosmic rays for different cosmic-ray components (light for the lower lines, heavy for the upper lines), based on the KASCADE-Grande (KG) data (\cite{KG:2013}, see text). The plain lines are the original fits of the KG collaboration. The dashed lines have been drawn to guide the eye. See also Fig.~\ref{fig:GCR2EGCRSketch} for a schematic idealization.}
\label{fig:globalPicture}
\end{center}
\end{figure*}

Qualitatively, as argued above, this suggests that the ankle does indeed mark the GCR/EGCR transition. Now, the recent progress in cosmic-ray studies has brought about additional information, which allows a more quantitative approach.

First:

\begin{enumerate}
\setcounter{enumi}{3}
\item the study of the UHECRs (notably their composition, see above) indicates that the EGCR component, which extends up to $10^{20}$~eV, dominates the overall spectrum in the most natural way only above $\sim 1\,\mathrm{to}\,3$~EeV.
\end{enumerate}

This is indeed consistent with a GCR/EGCR transition at the ankle.

Second:
\begin{enumerate}
\setcounter{enumi}{4}
\item the KASCADE-Grande data show a transition between a soft and a hard component of light nuclei (presumably dominated by protons) at $\sim 10^{17}$~eV (see the lowest line in Fig.~\ref{fig:globalPicture}).
\end{enumerate}

This ankle-like break in the light CR spectrum, if interpreted as a transition between the GCR protons and the EGCR protons, would imply that the energy spectrum of the Galactic Fe nuclei extends up to the ankle in the all-particle spectrum, which is indeed what is also suggested by the KASCADE-Grande data (see the upper lines in Fig.~\ref{fig:globalPicture}).

Finally, it might be added that:
\begin{enumerate}
\setcounter{enumi}{5}
\item the Auger results regarding the large scale anisotropies of the cosmic-rays around the ankle imply that, if Galactic, these nuclei must be heavy (typically Fe nuclei), in order not to produce an excessive dipole-like anisotropy.
\end{enumerate}

The whole set of available data is thus consistent with a rather natural and simple picture of both GCRs and EGCRs, which can be summarized as follows:
\begin{enumerate}[i.]
\item the ankle observed in the CR spectrum marks the transition from a dominantly Galactic component to a dominantly extragalactic component;
\item at this energy, the Galactic component is dominated by heavy nuclei (mostly Fe nuclei, presumably);
\item the GCR/EGCR transition for protons occurs at an energy of $\sim 10^{17}$~eV;
\item Galactic sources accelerate cosmic-rays up to an energy of $\sim Z\times 10^{17}$~eV (at least);
\item the EGCRs above the ankle exhibit a progressive transition towards heavier nuclei (probably reflecting the high-energy cutoff of the lighter nuclei at the source, as described within the framework of the so-called low proton $E_{\max}$ models).
\end{enumerate}

Note that, in this picture, the knee in the CR spectrum appears to be a break in the Galactic \emph{proton} spectrum, and the feature sometimes referred to as the ``second knee'' appears to be the corresponding break of the Galactic \emph{Fe component} (see Fig.~\ref{fig:globalPicture}). Such a distribution of breaks occurring for different nuclei at an energy proportional to $Z$ (i.e. at constant rigidity) is also consistent with the data collected by the KASCADE collaboration around the knee. From the theoretical point of view, such a break is indeed naturally expected, among other possibilities, as an effect of propagation of the CRs leaking out of the Galaxy (see \cite{Hoerandel:2004} for a review, and e.g. \cite{Giacinti+2014} for a recent account, with refs. therein).

We have added a few guiding lines and energy reference points to Fig.~\ref{fig:globalPicture}, to help following the global picture described above. This figure is from the KASCADE-Grande collaboration, and the plain lines and data points are theirs, showing the heavy component (say, Fe, for simplicity: top lines) and the light component (say, protons: bottom lines). We have added the dashed lines, prolongating the plain lines.

\begin{figure}[ht]
\begin{center}
\includegraphics[width=0.9\linewidth]{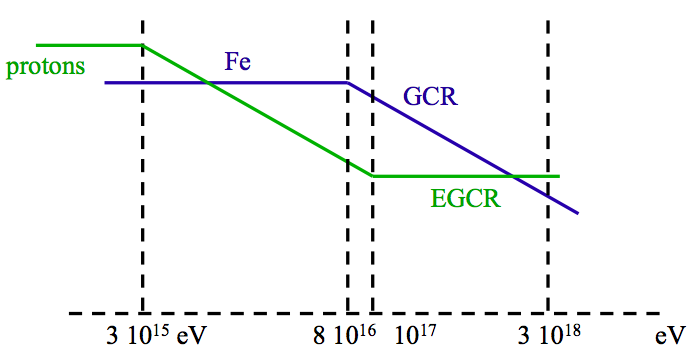}
\caption{Sketch of the GCR/EGCR transition, with the proton and Fe components indicated (respectively in green and in blue on the color version of the figure). In ordinate, the CR flux is multiplied by $E^{x}$, where $x$ is the logarithmic slope of the CR spectrum below the knee. (See also Fig.~\ref{fig:globalPicture}).}
\label{fig:GCR2EGCRSketch}
\end{center}
\end{figure}

Several points are interesting to note (see also Fig.~\ref{fig:GCR2EGCRSketch} for an schematic view of the global picture):
\begin{enumerate}[i.]
\item The position of the break in the heavy component is consistent with the Fe knee, i.e. the Fe counterpart of the global CR knee, occurring at $\sim 26$~times the energy of the proton knee.
\item The slope of the light component fitted by the KASCADE-Grande collaboration below the ankle-like break (in the light component) is the same as the slope fitted for the heavy component above the ``Fe-knee''. It is thus fully consistent with a picture in which whatever occurs to the protons, producing a steepening in the spectrum, occurs in exactly the same way to Fe nuclei, with a shift in energy by a factor $Z = 26$.
\item If one extends the pre-ankle (Galactic) bottom line to lower energies, one obtains a rather natural match to the measured flux at the knee. This gives further, quantitative support to the above picture.
\item If one extends the post-ankle bottom line (extragalactic in the view presented here) to higher energies, one obtains a rather natural match to the measured flux at the ankle, marking a transition between the heavy-dominated GCRs and the light-dominated EGCRs.
\end{enumerate}

Of course, one should remain cautious and not put too much weight on the exact levels of the different components. In particular, although the presence of an ankle in the lightest cosmic-rays at $\sim 10^{17}$~eV appears to be firmly established, the interpretation of the KASCADE-Grande data in terms of definite composition measurements depends on assumptions about the hadronic model used to describe the CR showers. However, we believe that the above (probably idealized) sketch is at least consistent with the current data, and offers a very simple and natural way to consider the different features observed in the spectrum as well as the transition from GCRs to EGCRs.

An additional comment should be made here. It has been objected that the Auger data on composition (related to both the average and the spread of the grammage corresponding to the maximum shower development in the atmosphere \cite{Auger:2010c}) are inconsistent with a picture in which heavy nuclei have a significant contribution to the flux in the ankle energy range. This is not correct. Although it is true indeed that the Auger data on composition around 1~EeV are consistent with cosmic rays being predominantly protons, it is also true that the data do not exclude, by no mean, a significant fraction of heavier nuclei, actually up to 50\%, as confirmed by explicit analyses developed within the Auger collaboration (taking into account a combination of protons, He, C, N, O, sub-Fe and Fe nuclei). In any case, at this stage, the presence of a roughly similar amount of protons and nuclei, whatever their origin, in the energy range of the ankle, cannot be excluded on the basis of the data. On the contrary, such a mixed composition is precisely what the KASCADE-Grande data suggest, as can be seen for instance in Fig.~\ref{fig:globalPicture}. If one were to argue that the Auger data exclude a non-negligible contribution of the heavy nuclei in the ankle energy range, then one would at best point a conflict between the Auger data and the KASCADE-Grande data, independently of any theoretical or phenomenological modeling. Our understanding of the data, however, based on discussions with involved members of both collaborations, is that no such conflict can be claimed today.

\section{The maximum energy problem of the standard SNR-GCR connection}

In the most consensual model for the origin of GCRs (which we refer to here as the SNR-GCR connection), the acceleration process is the so-called diffusive shock acceleration process  (or DSA, see e.g. \cite{JonesEllison1991,MalkovDrury2001} and refs. therein), and the acceleration of the cosmic rays occurs at the expanding forward shock of Galactic supernova remnants (SNRs), caused by the explosion of massive stars in the interstellar medium (as well as type-Ia supernov\ae). A number of prototype sources, including the so-called historic SNRs, have been extensively studied. A complete modeling of the multi-wavelenght emission of these sources has been attempted and often very convincingly achieved, including both thermal and non-thermal emissions, shock dynamics, particle acceleration, ionisation, heating, magnetic field amplification, back-reaction of the energetic particles on the shock structure, etc. This represents a remarkable success of the field of high-energy astrophysics.

The fact that particles not only \emph{can be}, but actually \emph{are} accelerated at the shock of SNRs, is a necessary condition for the SNRs to be considered the sources of the GCRs. It is however not sufficient. Another necessary condition is that the total power of the supernov\ae~in the Galaxy be sufficient to sustain the constant renewal of CRs. This condition appears to be met, indeed, and this is a key reason why the SNR-GCR connection has been so popular and so extensively studied. This condition is still not sufficient, however. If the SNRs are to be the sources of the GCRs, they must be able to accelerate particles according to a spectrum which, after propagation, matches the observed spectrum (and composition). This has long been recognized as a major challenge, notably because of the limitation imposed by the acceleration process itself to the maximum energy attained by the particles in the source.

\subsection{The problem}

It has been pointed out more than three decades ago that the diffusive shock acceleration process can hardly accelerate particles up to $10^{13}$~eV, assuming the most standard conditions in the SNR environment \cite{LagageCesarsky:1983}. This is more than two orders of magnitude below the knee. It was not considered as a show-stopper right away, however, because some hope remained that more detailed studies could push the limit higher and allow a significant fraction of the SNRs to ``reach the knee''. As a matter of fact, whether the maximum energy is limited by the acceleration time or by the size of the SNR, which must be (much) larger than the gyroradius of the particles that need to be confined in the vicinity of the shock, a key parameter is the magnetic field seen by the particles around the shock. Basically, both the particle gyroradii and the acceleration timescales are proportional to the amplitude of the magnetic field. Thus, if the latter is multiplied by 100, so will be the maximum energy too.

Such a large amplification of the ambient magnetic field around SNRs is indeed what has been suggested in the last decade, both on theoretical grounds, with resonant streaming instability as well as non-resonant, CR current driven instabilities (see e.g. the works of \cite{Bell:2004,DruryFalle:1986,BegelmanZweibel:1994,DownesDrury:2012,Bykov+2013,SchureBell:2013,Bykov+2014,DownesDrury:2014}), and on observational grounds, with the interpretation of the extreme thinness of the X-ray rims observed at the forward shock of SNRs as due to the synchrotron losses of the highest-energy electrons in the local, intense magnetic fields \cite{VinkLaming:2003,Berezhko+2003,Bamba+2004,Voelk+2005,Parizot+2006,Uchiyama+2007,Vink:2012}.

However, even with such a large field amplification, by a factor of order 100, and even if one assumes that this can be maintained at the appropriate length scale to be fully profitable to particle acceleration, it remains very difficult for DSA to accelerate protons up to 1~PeV in SNRs. Now, even if it were possible, this would still be very far from what is needed to validate, at least in principle, the SNR-GCR connection. Indeed, the above study of the GCR/EGCR transition, in the light of the experimental data on the highest energy cosmic-rays, shows that the GCR sources must be able to accelerate protons at least up to $\sim 10^{17}$~eV (and Fe nuclei up to the ankle). This is (at least) two orders of magnitude higher than what can be optimistically achieved by standard SNRs (note that, in most cases where a maximum proton energy could be estimated, this energy is actually much lower than 1~PeV, e.g. \cite{Parizot+2006}, even though this may simply be because the acceleration phase during which PeV energies are reached is very short in time).

\subsection{The ``other source'' solution}

One solution to save the SNR-GCR connection is to invoke an additional source of GCRs. Basically, the SNRs would account for the GCR spectrum up to the knee, while another type of sources would account for the spectrum above that. This other component could be either Galactic, or extragalactic (as investigated recently in \cite{Aloisio+2013}). This does not seem very likely, however, because the matching of the two independent components in a seamless knee-like structure is particularly difficult to achieve in practice (see above).

In the case of an additional Galactic component, one may also refer to Ockham's razor and note that, if another type of sources, still to be identified, is needed to explain the cosmic-ray flux up to the ankle, then there might be no need for SNRs at all (at least not as the main contributors to the GCRs), as these new sources might just as well dominate at lower energy too, instead of suddenly starting above the PeV energy range. This would solve the problem of matching two components through a knee.

\subsection{The ``subset'' solution}

Another interesting solution to save the SNR-GCR connection consists in invoking a subset of the SNRs, rather than a new type of sources, to fill the gap between the knee and the ankle. This is an attractive solution because there could then be no need to match two independent components in a seamless way. If most SNRs accelerate particles up to the knee, and a smaller and smaller subset manages to accelerate particles up to higher and higher energies (eventually reaching $\sim Z\times 10^{17}$~eV or above, the schematic picture presented above would be globally satisfied. The knee would then be the consequence of the reduction of the number of contributing sources rather than an effect of particle propagation in (and escape out of) the Galaxy.


For instance, the flux of GCRs above the knee could be due some exceptional SN explosions or failed gamma-ray bursts, or to SNRs which interact with the wind of the progenitor (massive) star during the first few years after the explosion (\cite{VoelkBiermann1988,Biermann1993,Tatischeff2009}).

Of course, it remains to be confirmed that this ``subset solution'' is viable in practice, i.e. that some SNRs can indeed accelerate particles up to $Z\times 10^{17}$~eV. Such a demonstration would be a remarkable achievement, and it certainly is an important astrophysical question. According to \cite{Tatischeff2009}, in the case of SN~1993J, the proton maximum energy could have been as high as 2--3~$10^{16}$~eV, one or two days after the outburst. Although this is still short of what is needed, and although, according to \cite{Tatischeff2009}, it does not take into account some nonlinear effects pointed out by \cite{EllisonVladimirov2008} which could reduce further the maximum energy, it suggests that the maximum GCR energies could be reached in a few exceptional sources, in some particular circumstances, so that the usual limitations pertaining to the most studied SNRs (in a later phase of their evolution) may not be directly relevant.

However, while this remains an open question, we wish to point out that the subset of sources needed to account for the whole GCR spectrum should not be a small subset, as discussed below.

\begin{figure}[t!]
\begin{center}
\includegraphics[width=\linewidth]{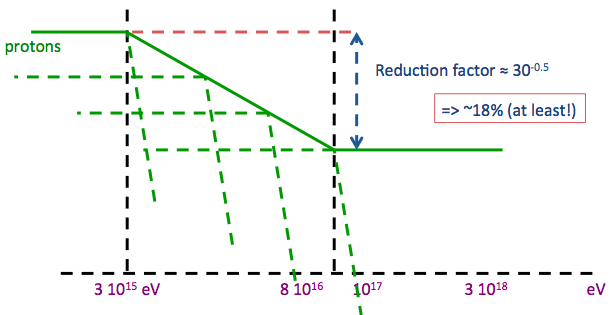}
\caption{Sketch of the GCR proton flux above the knee, showing the contribution of a smaller and smaller number of sources at higher and higher energy, up to $\sim 10^{17}$~eV, where the EGCR proton component becomes more abundant (compare with the ``proton line'' of Fig.~\ref{fig:GCR2EGCRSketch}). The various dashed lines show, schematically, the contributions of all the sources which contribute up to a given energy (where the dashed line touches the plain line), with an arbitrary cut-off above that energy.}
\label{fig:fractionOfSources}
\end{center}
\end{figure}

\subsection{The size of the subset in the ``subset solution''}

Let us adopt the point of view that all SNRs accelerate protons with a power-law in $E^{-x}$ up to the knee, at $E_{\mathrm{knee}} = 3\times 10^{15}$~eV, or above, while only a smaller fraction of SNRs, $\alpha(E)$, accelerate protons with the same power-law up to an energy $E > E_{\mathrm{knee}}$ or above. What is then the fraction of the SNRs, $\alpha(10^{17}\,\mathrm{eV})$, which must accelerate protons up to $10^{17}$~eV or above?

One can find a lower limit to this fraction, by determining the ``flux deficit'', compared to the reference case where all sources would contribute all the way to $10^{17}$~eV, without any spectral break. This is illustrated by the schematic view of Fig.~\ref{fig:fractionOfSources}, where the plain line is the same as the bottom broken line in Fig.~\ref{fig:GCR2EGCRSketch} (sketch of the proton component), and each dashed line represents the contribution of all the sources which contribute up to a given energy, where the dashed line touches the plain line (with an arbitrary cut-off above that energy). The change of slope across the knee, as obtained from the data, corresponds roughly to a change of index of $0.5$. With a lever arm of $\sim 30$ in energy, this represents a flux reduction factor (compared to the extrapolated value, shown by the red dashed line) of the order of $30^{-0.5}\sim 0.18$. In other words, the fraction $\alpha(10^{17}\,\mathrm{eV})$ of exceptional GCR sources which must be invoked in the framework of the ``subset solution'' is of the order of 15--20\%, at least.

It is interesting to note that this estimate was made possible by the KASCADE-Grande data, which fix a point of reference for the proton component. Although the corresponding flux remains incertain, because of the uncertainty in the high-energy hadronic models, the cut in the data (to select the lightest showers) is presumably conservative, and the proton fraction is more likely to be somewhat larger than lower (M. Bertaina, private communication: the very existence of an ankle-like break in the data shows that there is not too much ``pollution'' by heavier nuclei).

The reason why we indicated that the above fraction is a lower limit is that this estimate does not take into account any other possible reduction factor in the comparison between the proton flux at $\sim 10^{17}$~eV and the extrapolation of the proton flux below the knee. In particular, we assumed that there were no reduction due to a loss of confinement of the GCRs, or, more precisely, a change in the energy-dependence of the rate of leakage. This is most probably an ``optimistic'' view, and some models, on the contrary, estimate that the knee can be fully explained by the energy-dependent escape out of the Galaxy (see e.g.~ref.~\cite{Giacinti+2014} for a recent discussion). If the magnetic field assumed by these authors, with a coherence length of the turbulent component of the order of 5~pc, is confirmed to represent the Galactic magnetic field reasonably well, then the GCR source spectrum should show no break at all at the knee. In this case, 100\% of the sources should accelerate protons up to $10^{17}$~eV.

In conclusion, depending on the actual level of the propagation/leakage effect, the ``subset solution'' requires a fraction of a least $\sim 20\%$, and possibly up to possibly 100\% of the SNRs to accelerate protons up to $10^{17}$~eV. This seems difficult in the current stage of knowledge, but future studies might demonstrate that this is nevertheless the case.

\section{A few more elements}

We have considered above some aspects of the SNR-GCR connection scenario, in the light of the recent data on cosmic rays above the knee(s), around the ankle(s) and up to the ultra-high-energy domain. The main result of these considerations is that the GCR sources should accelerate particles up to $\simeq Z\times 10^{17}$~eV (or more), which seems to be a serious challenge for the diffusive shock acceleration process operating in the environment of standard, isolated SNRs. In this section, we briefly review some additional elements of the cosmic-ray phenomenology, which may be relevant to the general problem of the GCR origin.

\subsection{Light element nucleosynthesis}

A very important aspect of the low-energy CR phenomenology is related to the nuclear reactions suffered by the energetic particles, while they propagate from their sources to the Earth. These reactions result in a modification of their elemental and isotopic composition: the primary cosmic-rays, injected by the sources in the interstellar medium (ISM), can be destroyed by spallative reactions with the ambiant material and see their abundance decrease, while secondary particles can be produced in flight by these very reactions, and see their abundance increase. It is well known that the secondary-to-primary ratios can be used to derive the amount of material (or grammage, in g/cm$^{2}$) that the GCRs have gone through, as a function of energy. In the case of radioactive secondary nuclei, also called \emph{cosmic-ray clocks}, the secondary-to-primary ratios are also used to derive the confinement time of the GCRs.

Among the secondary nuclei, the so-called light elements, Li, Be and B (or LiBeB, in short), are particularly important, because they are almost exclusively produced by the GCR-induced spallation reaction in the ISM (this is true at least for the $^{6}$Li, $^{9}$Be and $^{10}$B isotopes). The main reactions are spallation reactions involving C, N and O nuclei (or CNO, in short). One can distinguish between two channels of this so-called spallative nucleosynthesis: the \emph{direct spallation}, where protons and He nuclei among the GCRs break the CNO in the ISM, and produce LiBeB essentially at rest in the ISM, and the \emph{inverse spallation}, where CNO nuclei among the GCRs are broken in flight into LiBeB nuclei, which thus remain part of the GCRs.

The LiBeB production rate in the Galaxy can be obtained by multiplying the appropriate energy average of the CR flux by the spallation cross section and the target (ISM) density, and integrating over space. It has been known for more than four decades that the correct amount of LiBeB in the Galaxy is obtained when integrating this production rate over the age of the Galaxy \cite{Reeves+1970,MAR1971}, which validates the general concept of GCR-induced spallative nucleosynthesis.

In the 1990's, however, a crisis appeared in this beautiful picture, following the measurement of the LiBeB abundace is the atmosphere of very low-metallicity stars in the Galaxy. According to the standard GCR scenario, the LiBeB isotopes are the perfect prototypes of secondary isotopes in the chemical evolution of the Galaxy. Indeed, to synthesize a LiBeB nucleus, one first needs to have CNO nuclei to break. The LiBeB production rate is directly proportional to the CNO abundance in the ISM (for the direct spallation) or in the GCRs (for the inverse spallation). In the very early Galaxy, say at the first generation of stars, there were essentially no CNO in the ISM, so the LiBeB production rate was essentially zero. Likewise, the shocks of the first generation of supernov\ae~had no CNO to accelerate, so the LiBeB production rate by reverse spallation was also essentially zero. The efficiency of the process then increased with time (or ``chemical time''), depending directly on the ambiant metallicity.

Let $Z(t)$ be the metallicity of the Galaxy at time $t$ of its chemical evolution (say, the abundance of CNO, for simplicity, which is a good tracer in the early Galaxy). It is roughly proportional to the total number of supernova\ae, $\mathrm{SN}(t)$, which have exploded up to time $t$ (especially in the early Galaxy, where type Ia SNe do not play an important role). In the standard SNR-GCR connection scenario, the flux of GCRs at time $t$ is proportional to the SN rate, $\mathrm{d}[\mathrm{SN}(t)]/\mathrm{d}t$. Then, the LiBeB production rate $\mathrm{d}[\mathrm{LiBeB}(t)]/\mathrm{d}t$, whether by direct or inverse spallation, is proportional to $Z(t)$ (whether the target or the source metallicity) and to the GCR flux, i.e. $\mathrm{d}[\mathrm{SN}(t)]\mathrm{d}t$. Finally, one finds that $\mathrm{d}[\mathrm{LiBeB}(t)]/\mathrm{d}t\propto Z\mathrm{d}Z/\mathrm{d}t$, which integrates into $\mathrm{LiBeB}(Z)\propto Z^{2}$.

This is the standard behavior for a secondary element: as the Galactic chemical evolution proceeds, its abundance grows proportionally to the square of the metalliticity, or, if one prefers, its abundance ratio to the abundance of O, say, is proportional to the metallicity. Now, what has been observed is that the ratio of the LiBeB abundance to the O abundance is actually constant (see \cite{VangioniFlam+1998} for a review, and refs. therein \cite{Gilmore+1992,Ryan+1992,BoesgaardKing1993,Duncan+1992,Ryan+1996,Walker+1993,Ramaty+1997}). In other words, the LiBeB elements behave as primary elements, not secondary ones! This is in direct contradiction with the prediction of the standard SNR-GCR connection scenario.

In order to reproduce this primary behavior, LiBeB production models must ensure that the production rate is independent of the preexisting amount of CNO in the ISM. This is definitely impossible if the sources of the GCRs are SNRs propagating in the standard ISM, accelerating the ambiant material.

A possibility could be that the SNe manage to accelerate their own ejecta. Unfortunately, this is strongly believed not to be the case, because of the inefficiency of the reverse shock, and because the fast moving ejecta which could exceptionally cross the forward shock of the SNR (e.g.~through hydrodynamical instabilities) would remain insignificant. The other possibility could be that the particles accelerated at the forward shock and advected downstream towards the interior of the SNR could interact with the CNO of the SN ejecta, if sufficient mixing of the material occurs and the energetic particles can diffuse to a region where a significant abundance of CNO is present. Unfortunately again, detailed analyses show that such a process cannot be efficient enough \cite{ParizotDrury1999a,ParizotDrury1999b} (Tatischeff, private communication).

In conclusion, the standard SNR-GCR scenario has been shown to be inconsistent with the observational data on LiBeB chemical evolution. This is a serious problem which should be considered attentively when investigating the GCR origin problem.

However, the LiBeB evolution problem has a well-known solution, through the so-called superbubble model, which is relevant also in the light of the remarks below.

\subsection{Superbubbles vs. isolated SNRs}

\subsubsection{SN power}

One of the main arguments in favor of the standard SNR-GCR connection is that the power needed to maintain the GCR flux throughout the Galaxy is around 10\%--20\% of the total kinetic power of the SN shock waves, and that this fraction is compatible with the expected efficiency of the diffusive shock acceleration (DSA) process in SNRs.

However, most of the studies of DSA concentrate on \emph{isolated} SNRs, corresponding to SNe exploding in an environment representative of the general ISM, possibly with some density gradient, but usually smooth on the scale of the shock. In the modeling of particular SNRs, the interaction of the shock with dense clouds can and sometimes must be taken into account, but such interactions are usually take place in a given region, and the main effect is a local modification of the density of the ambiant medium and of the shock parameters.

It is well known, however, that most SNe are not isolated from one another. The vast majority of the massive stars in the Galaxy are born in groups called OB associations, in star formation episodes which are localized in both space and time. Because the massive stars are short lived, they do not have time to wander far away from their siblings before they enter the final stages of their evolution and eventually explode as supernov\ae. As a consequence, OB associations give rise to repeated SN explosions, also preceded by powerful massive stellar winds, injecting a very large amount of energy in a relatively small region of space. This energy input, by both stellar winds and SN explosions, results in the formation and growth of vast structures, much larger than SNRs, referred to as superbublles. The basic properties and evolution of these superbubbles, which are observed in large numbers in the Milky-Way and nearby galaxies, have been studied and modeled since decades (e.g. \cite{Castor+1975,Weaver+1977,MacLowMcCray1988,MacLow+1989,ChuMacLow1990,Ferriere+1991,BaumBreit2013}).

From the point of view of the GCR origin, the main consequence of this fact is that most of the SN power injected in the Galaxy is actually not released in the ISM through isolated SN events, but inside superbubbles. Overall, probably $90\pm10$\% of the core-collapse supernov\ae, and $\sim 75$\% of all Galactic supernov\ae occur inside a superbubble (e.g. \cite{Higdon+1998,HigdonLingenfelter2005}).

\subsubsection{Particle acceleration and maximum energy}

A natural question to ask, then, is whether this does or does not make a difference regarding particle acceleration. This question has been addressed in a systematic way as early as in the 1990's, notably by Bykov and collaborators \cite{BykovFleishman1992,Bykov+1995,BykovToptygin2001,Bykov2001}. It was further shown that collective acceleration effects (rather than a series of individual and independent acceleration events, by successive supernova\ae) should indeed be expected in the environment of OB associations \cite{Parizot+2004}. The specificity of particle acceleration inside superbubbles is due to the presence of an ensemble of shocks with a distribution of Mach numbers, produced by the interaction of stellar winds, SN shocks and reflected shocks off denser clumps, in an otherwise hot, tenuous and presumably highly magnetized medium, where an intense supersonic turbulence is expected to be found.

Theoretical studies of particle accelerationÊinside a superbubble (see \cite{Bykov2001} and refs. therein) suggest that acceleration could indeed be very efficient, with a large power conversion factor on average (although the flux of energetic particles is probably subject to strong intermittency), which could also help solving the so-called superbubble energy crisis \cite{ButtBykov2008}. The resulting energy spectrum is also time-dependent, but once integrated over time, it is likely to be a power-law with a rather steep index, compatible with the inferred GCR source spectrum \cite{Bykov2001}.

Admittedly, this conclusion is still preliminary, since the typical parameters of a real-life superbubble are not well known, and little observations can be called upon to verify the main predictions of the models. This is mostly due to the large angular scale of the nearby superbubbles and the diffuse nature of the associated non thermal emission, which makes decisive X-ray and gamma-ray observations very difficult, with predicted flux below the current detection capabilities. However, a key aspect of particle acceleration inside superbubbles is that it does not suffer from the same limitation as isolated SNRs regarding the maximum energy. Simple order-of-magnitude estimates, confirmed by more detailed numerical calculations, show that superbubbles should indeed be able to accelerate particles up to $Z\times 10^{17}$~eV \cite{Bykov+1995,Bykov2001}.

To date, there is no complete and definite alternative model for the origin of GCRs based on superbubbles, notably because these environments are still poorly known from the observational point of view and because the number of actual theoretical studies of particle acceleration inside superbubbles is very limited. However, we believe that, if SNe are indeed to be the main power source of GCRs, then superbubbles \emph{must} be taken into account seriously in any GCR source model, because this is the place where most of this power is released.

\subsubsection{LiBeB evolution}

Besides, as anticipated above, superbubbles offer the only known solution to the LiBeB evolution problem. The way this solution works is as simple as inevitable: when supernov\ae~explode repeatedly inside a superbubble, the ambiant medium swept up by a given supernova shock is enriched in the ejecta of the previous SNe and massive stellar winds. In this way, although it is essentially impossible for a supernova to accelerate efficiently its own ejecta and thus produce LiBeB isotopes with high efficiency in a very metal-poor environment, a superbubble, as a whole, can easily accelerate the ejecta of several tens of SNe whose progenitors belong to the same OB association, before they disperse and mix with the ambiant medium. In such an environment, the metallicity of the accelerated particles does not depend strongly on the preexisting metallicity of the ISM (especially in the very early Galaxy, where the LiBeB problem is most acute), and the GCR-induced LiBeB production efficiency is essentially independent of time, as required by the data.

Detailed quantitative studies have shown that this simple idea does indeed meet the observational requirements in a very natural way. For more details about how this works in practice, see \cite{ParizotDrury1999,ParizotDrury2000a,ParizotDrury2000b,Parizot2001,Higdon+1998,HigdonLingenfelter2005}.

It is not clear to us why the LiBeB evolution problem, where an observational fact appears to be in direct contraction with the standard SNR-GCR connection scenario (supported by many observational and theoretical studies over nearly two decades and for which a consensus has been reached in the community) is not taken into account in a more systematic way in the discussions about the origin of cosmic rays.

\subsubsection{Isotopic anomalies}
\label{sec:Isotopic}

Another important clue about the cosmic-ray origin is the GCR source composition, which can be inferred from the measured composition, after an appropriate account of the nuclear reactions taking place during the particle transport to the Earth. 

For instance, it is known since the 1970's that the GCR source composition is particularly rich in $^{22}$Ne \cite{Maehl+1975,GarciaMunoz+1979,Duvernois+1996,Webber+1997}, the $^{22}$Ne/$^{20}$Ne isotopic ratio being $\sim 5$ times larger in the GCR source composition than in the solar wind (\cite{Binns+2006a,Binns+2006b} and references therein). This isotopic ratio is definitely larger than would be expected if isolated SNRs were the main GCR sources. If, on the contrary, GCRs originate mostly from superbubbles, then the $^{22}$Ne excess can be easily understood as resulting from the pre-enrichment of the accelerated material by the ejecta of massive stars, since $^{22}$Ne is essentially released in the ISM by the winds of Wolf-Rayet stars (see \cite{HigdonLingenfelter2003,HigdonLingenfelter2006,LingenfelterHigdon2007,Binns+2006b} for more detail).

Note that $^{22}$Ne is not the only isotope pointing towards an enrichment of the GCR source material by massive star's ejecta. As noted by \cite{Binns+2006b}, the three largest GCR isotopic anomalies predicted by Wolf-Rayet models, $^{12}$C/$^{16}$O, $^{22}$Ne/$^{20}$Ne and $^{58}$Fe/$^{56}$Fe, are present in the GCRs. The authors conclude that superbubbles are the likely source of, at least, a substantial fraction of the GCRs. Likewise, \cite{HigdonLingenfelter2005} argue that $\sim 88\%$ of the cosmic-ray heavy particles are accelerated inside superbubbles.

At this point, it isÊalso worth mentioning that extended observational capability in the MeV range would be very valuable to confront models of GCR origin with \emph{in situ} data on the composition of the accelerated particles. Many important nuclear de-excitation lines lie in this energy range. If sensitive enough instruments could detect such $\gamma$-ray lines inside or close to a given acceleration site, one could directly confront the energetic particle composition with the required composition of the GCRs at their source, and even access isotopic ratios. These ratios are particularly discriminative in this perspective, because they are not sensitive to any of the potentially complex chemical processes which may play a role in the way different particles are injected into the acceleration process (e.g. \cite{Meyer+1997}). Theoretical estimates and predictions regarding such nuclear de-excitation lines have been discussed in a general context for several decades \cite{Ramaty+1979a,Ramaty+1979b}, and also explicitly in the context of constraining the origin of cosmic-rays \cite{Tibolla+2011a}, underlying the importance of project missions such as GRISP \cite{Tibolla+2011b}. A similar point, connecting energetic particles and nuclear de-excitation lines, had also been emphasized in the context of the above-mentioned LiBeB problem \cite{Casse+1995a}.

\subsubsection{Distribution of GCRs across the Galaxy}

Finally, we briefly mention that a superbubble origin of (at least some of) the GCRs would also have some consequences on their transport across the Galaxy. It is well known that the distribution of GCRs shows very little anisotropy at the Earth, even at high energy, and that their flux appears quite homogeneous throughout the Galaxy. This last point is not so easy to reconcile with the idea that the GCR sources are individual SNRs, with a distribution similar to that of the molecular gas in the disk of the Galaxy. A much larger concentration of sources would then be expected in the so-called molecular ring, at a galactocentric radius $r \simeq 5$~kpc, which should somehow reflect in the distribution of the GCR density. This problem may not be so severe, and solutions may be proposed, notably by assuming a very large confinement ``halo'' extending far away from the Galactic disk (beyond 5~kpc).

However, it is interesting to note that if most GCRs are initially accelerated inside superbubbles, their effective injection in the general ISM should be quite different from what would be otherwise expected. Indeed, superbubbles are known to eventually ``breakout'' the Galactic disk, as they grow asymmetrically in the direction opposite to the density gradient, away from the central plane (see e.g. \cite{MacLow+1989,BaumBreit2013,Roy+2013}). As the superbubbles blow out of the disk, a substantial fraction of the energetic particles in their interior should be driven directly into the halo, where they can diffuse more efficiently and ``rain down'' back to the disk in a more homogeneous way.

We believe that such a phenomenon should be kept in mind when addressing the general question of GCR transport in the Galaxy.

\section{Conclusion}

In this paper, we have examined the question of the Galactic cosmic ray origin in an enlarged context, taking into account the data available at ultra-high-energy, and paying particular attention to the transition between the Galactic and the extragalactic components of cosmic rays, in the light of the recent KASCADE-Grande data.

We have shown that the data strongly suggest that (and can be understood in a very natural way if) the overall GCR/EGCR transition occurs at the ankle, around $3\times 10^{18}$~eV, while the extragalactic protons start to dominate over Galactic protons above $\sim 10^{17}$~eV.

An important consequence is that the GCR sources must be able to accelerate protons with a relatively smooth spectrum (at least) up to $10^{17}$~eV. This is a serious problem for the standard SNR-GCR connection scenario. While the viability of this standard scenario could be saved if a subset of SNRs could be demonstrated to achieve such high energies, we have shown that this subset of sources should not be small. The ``exceptional'' accelerators should represent at least 20\% of all the source (in terms of power), assuming that particle propagation in the Galactic magnetic field plays no role at all in shaping the knee in the GCR spectrum, or up to 100\% if the knee is entirely due to such effects. In the latter case, the GCR source spectrum could thus a be a simple power-law up to the highest energies achieved in the Galactic accelerators -- a rather attractive possibility, from the point of view of simplicity.

Given the importance of UHECRs to draw a global picture of the cosmic ray phenomenon over the whole energy range, we note that future extensions and/or enhancements of the Telescope Array and Auger experiments could provide interesting new data in the coming years. Likewise, at the highest energy end of the spectrum, the JEM-EUSO space mission should mark a decisive step towards the long sought cosmic-ray astronomy, thanks to its unprecedented aperture \cite{BertainaParizot:2014}.

We have then turned to a very important aspect of the GCR phenomenology, namely the nucleosynthesis and Galactic evolution of the light elements: Li, Be and B. We recalled that the standard SNR-GCR connection scenario fails to account for the primary behavior of the LiBeB in the early Galaxy, and that isolated supernov\ae~simply cannot have been the main sources of the GCRs in the past. The consensus concerning the LiBeB evolution problem is that most cosmic rays have been (and thus probably still are) accelerated inside superbubbles, in an environment which is highly enriched in freshly synthesized material released by the winds of massive stars and SN explosions.

We recalled that this is anyway a rather natural expectation, given that the main part of the SN power in the Galaxy is released inside superbubbles.

In addition, we recalled that the superbubble origin of most of the GCRs has also been invoked in a different context, to solve some persistent problems related to isotopic anomalies in the source composition of GCRs. Since isotopic ratios are not subject to selection effects based on chemistry (like the refractory/volatile segregation, \cite{Meyer+1997}), these data are important clues which should be considered with great attention when tackling the GCR origin puzzle.

It might be argued that whether the acceleration of most GCRs takes place in superbubbles or not does not change much the standard SNR-GCR connection after all, since the latter scenario does not explicitly mention in which environment the SNR shock actually propagates. However, theoretical studies strongly suggest that particle acceleration processes inside superbubbles should be significantly different from the standard diffusive shock acceleration mechanism at the well-defined shocks of individual SNRs. Not only does a SNR shock propagate differently in the hot, tenuous and turbulent medium of a superbubble, but collective effects are expected to be important \cite{Bykov2001,Parizot+2006}, and the interaction of multiple primary and secondary shocks should lead to the establishment of a complex supersonic magnetic turbulence, and a specific acceleration mechanism (see references above).

Another important specificity of superbubbles, of course, is that they do not appear to suffer from the same limitations as the individual SNRs regarding the maximum energy. If GCR sources do indeed have to accelerate particles up to $\sim Z\times 10^{17}$~eV, as argued above, superbubbles should be a natural candidate.

It should be reminded that there is not yet a fully established model of particle acceleration inside superbubbles. Important works have been published in this framework over the years, but a more dedicated interest of the community would be needed to explore this possibility and see whether a complete model of the GCR origin could be designed in this framework and tested in a more direct way. This will also require increased observational capability, e.g. in the MeV range, where nuclear de-excitation lines could be detected. In this respect, as well as for direct energetic particle composition measurements (cf. Sect.~\ref{sec:Isotopic}), project missions such as GRIPS \cite{Tibolla+2011a} or the Compton Cube \cite{Lebrun+2003} would be extremely valuable.

Overall, we hope that this paper can contribute to a general effort to open a larger perspective in the search for the origin of cosmic-rays, which goes beyond the standard SNR-GCR connection and takes into account a wide range of data, including GCR source composition, spectrum, homogeneity and anisotropy, as well as the data recently collected in the energy range of the ankle and on ultra-high-energy cosmic rays. This may also release some pressure on the modeling of SNRs. It is our feeling that, for historical reasons, a number of authors interested in the modeling of the multi-wavelength emission of individual SNRs feel the need that their models, often very successful indeed, account \emph{in addition} for the various aspects of the GCR phenomenology. This may not be required if the kind of individual SNRs they are mostly interested in are actually not the main sources of cosmic-rays\ldots

\section*{Acknowledgements}
I would like to warmly thank Denis Allard for continuous discussions about cosmic rays over more than a decade, as well as Mario Bertaina for detailed explanations about the KASCADE-Grande results and analysis. I wish to express my gratitude to Luke Drury and Omar Tibolla for organizing a very interesting and enriching Workshop on cosmic-ray origin in the exceptional environment of San Vito di Cadore. Special thanks to Omar Tibolla for his remarkable patience and efficiency.




\nocite{*}
\bibliographystyle{elsarticle-num}
\bibliography{CRBTSM_parizot}

\begin{thebibliography}{10}
\expandafter\ifx\csname url\endcsname\relax
  \def\url#1{\texttt{#1}}\fi
\expandafter\ifx\csname urlprefix\endcsname\relax\def\urlprefix{URL }\fi
\expandafter\ifx\csname href\endcsname\relax
  \def\href#1#2{#2} \def\path#1{#1}\fi

\bibitem{Greisen1966}
K.~Greisen, {End to the Cosmic-Ray Spectrum?}, Phys. Rev. Lett. 16 (1966) 748.

\bibitem{ZatsepinKuzmin1966}
G.~T. Zatsepin, V.~A. Kuz'min, {Upper Limit of the Spectrum of Cosmic Rays},
  Sov. Phys. JETP Lett. 4 (1966) 78.

\bibitem{Allard:2012}
D.~Allard, {Extragalactic propagation of ultrahigh energy cosmic-rays},
  Astropart. Phys. 39 (2012) 33--43.

\bibitem{Harari+2006}
D.~{Harari}, S.~{Mollerach}, E.~{Roulet}, {On the ultrahigh energy cosmic ray
  horizon}, JCAP 11 (2006) 12.
\newblock \href {http://arxiv.org/abs/astro-ph/0609294}
  {\path{arXiv:astro-ph/0609294}}.

\bibitem{Globus+2008}
N.~{Globus}, D.~{Allard}, E.~{Parizot}, {Propagation of high-energy cosmic rays
  in extragalactic turbulent magnetic fields: resulting energy spectrum and
  composition}, A\&A 479 (2008) 97--110.
\newblock \href {http://arxiv.org/abs/0709.1541} {\path{arXiv:0709.1541}}.

\bibitem{Kachelriess+2009}
M.~{Kachelrie{\ss}}, E.~{Parizot}, D.~V. {Semikoz}, {The GZK horizon and
  constraints on the cosmic ray source spectrum from observations in the GZK
  regime}, Soviet Journal of Experimental and Theoretical Physics Letters 88
  (2009) 553--557.
\newblock \href {http://arxiv.org/abs/0711.3635} {\path{arXiv:0711.3635}}.

\bibitem{DeDomenicoInsolia2013}
M.~{De Domenico}, A.~{Insolia}, {Influence of cosmological models on the GZK
  horizon of ultrahigh energy protons}, Journal of Physics G Nuclear Physics
  40~(1) (2013) 015201.
\newblock \href {http://arxiv.org/abs/1104.5083} {\path{arXiv:1104.5083}}.

\bibitem{Blaksley+2013}
C.~Blaksley, E.~Parizot, G.~Decerprit, D.~Allard, {Ultra-high-energy cosmic ray
  source statistics in the GZK energy range}, A{\&}A 552 (2013) 125.

\bibitem{KoteraOlinto2011}
K.~Kotera, A.~V. Olinto, {The Astrophysics of Ultrahigh-Energy Cosmic Rays},
  ARA{\&}A 49 (2011) 119.

\bibitem{HiRes:2008a}
{High Resolution Fly's Eye Collaboration}, {First Observation of the
  Greisen-Zatsepin-Kuzmin Suppression}, Phys. Rev. Lett. 100 (2008) 101101.

\bibitem{Auger:2008a}
{Pierre Auger Collaboration}, {Observation of the Suppression of the Flux of
  Cosmic Rays above $4\,10^{19}$~eV}, Phys. Rev. Lett. 101~(6) (2008) 61101.

\bibitem{Auger:2010a}
{Pierre Auger Collaboration}, {Measurement of the energy spectrum of cosmic
  rays above $10^{18}$~eV using the Pierre Auger Observatory}, Phys. Lett. B
  685~(4) (2010) 239--246.

\bibitem{TA:2013a}
T.~Abu-Zayyad, R.~Aida, M.~Allen, et~al., {The Cosmic-Ray Energy Spectrum
  Observed with the Surface Detector of the Telescope Array Experiment}, ApJL
  768~(1) (2013) L1.

\bibitem{Auger:2010c}
{Pierre Auger Collaboration}, {Measurement of the Depth of Maximum of Extensive
  Air Showers above 10$^{18}$~eV}, Phys. Rev. Lett. 104~(9) (2010) 91101.

\bibitem{Allard+2008}
D.~Allard, N.~G. Busca, G.~Decerprit, A.~V. Olinto, E.~Parizot, {Implications
  of the cosmic ray spectrum for the mass composition at the highest energies},
  JCAP 10~(1) (2008) 033.

\bibitem{Aloisio+2011}
R.~{Aloisio}, V.~{Berezinsky}, A.~{Gazizov}, {Ultra high energy cosmic rays:
  The disappointing model}, Astroparticle Physics 34 (2011) 620--626.
\newblock \href {http://arxiv.org/abs/0907.5194} {\path{arXiv:0907.5194}}.

\bibitem{TAHotSpot:2014}
R.~U. {Abbasi}, M.~{Abe}, T.~{Abu-Zayyad}, {et al.}, {Indications of
  Intermediate-scale Anisotropy of Cosmic Rays with Energy Greater Than 57~EeV
  in the Northern Sky Measured with the Surface Detector of the Telescope Array
  Experiment}, Astrophys. Journal Lett. 790 (2014) L21.

\bibitem{Auger:2007a}
{Pierre Auger Collaboration}, {Correlation of the Highest-Energy Cosmic Rays
  with Nearby Extragalactic Objects}, Science 318~(5) (2007) 938--.

\bibitem{Auger:2010b}
{Pierre Auger Collaboration}, {Update on the correlation of the highest energy
  cosmic rays with nearby extragalactic matter}, Astropart. Phys. 34~(5) (2010)
  314--326.

\bibitem{RouilleDOrfeuil+2014}
B.~{Rouill{\'e} d'Orfeuil}, D.~{Allard}, C.~{Lachaud}, E.~{Parizot},
  C.~{Blaksley}, S.~{Nagataki}, {Anisotropy expectations for ultra-high-energy
  cosmic rays with future high-statistics experiments}, Astronomy and
  Astrophysics 567 (2014) A81.
\newblock \href {http://arxiv.org/abs/1401.1119} {\path{arXiv:1401.1119}}.

\bibitem{Aloisio+2013}
R.~{Aloisio}, V.~{Berezinsky}, P.~{Blasi}, {Ultra high energy cosmic rays:
  implications of Auger data for source spectra and chemical composition},
  ArXiv e-prints\href {http://arxiv.org/abs/1312.7459}
  {\path{arXiv:1312.7459}}.

\bibitem{Allard+2005}
D.~Allard, E.~Parizot, A.~V. Olinto, E.~Khan, S.~Goriely, {UHE nuclei
  propagation and the interpretation of the ankle in the cosmic-ray spectrum},
  A{\&}A 443~(3) (2005) L29--L32.

\bibitem{Allard+2007}
D.~Allard, E.~Parizot, A.~V. Olinto, {On the transition from galactic to
  extragalactic cosmic-rays: Spectral and composition features from two
  opposite scenarios}, Astropart. Phys. 27~(1) (2007) 61--75.

\bibitem{KG:2011}
W.~D. {Apel}, J.~C. {Arteaga-Vel{\'a}zquez}, K.~{Bekk}, {et al.}, {Kneelike
  Structure in the Spectrum of the Heavy Component of Cosmic Rays Observed with
  KASCADE-Grande}, Physical Review Letters 107~(17) (2011) 171104.
\newblock \href {http://arxiv.org/abs/1107.5885} {\path{arXiv:1107.5885}}.

\bibitem{KG:2013}
W.~D. {Apel}, J.~C. {Arteaga-Vel{\'a}zquez}, K.~{Bekk}, {et al.},
  {KASCADE-Grande measurements of energy spectra for elemental groups of cosmic
  rays}, Astroparticle Physics 47 (2013) 54--66.

\bibitem{Bertaina:2014}
M.~{Bertaina}, W.~D. {Apel}, J.~C. {Arteaga-Vel{\'a}zquez}, {et al.}, {The
  cosmic ray spectrum and composition measured by KASCADE-Grande between
  $10^{16}$~eV and $10^{18}$~eV}, in: O.~{Tibolla}, L.~{Drury} (Eds.), Cosmic
  Ray Origin beyond the Standard Model, this volume, 2014.

\bibitem{Hoerandel:2004}
J.~R. {H{\"o}randel}, {Models of the knee in the energy spectrum of cosmic
  rays}, Astroparticle Physics 21 (2004) 241--265.
\newblock \href {http://arxiv.org/abs/astro-ph/0402356}
  {\path{arXiv:astro-ph/0402356}}.

\bibitem{Giacinti+2014}
G.~{Giacinti}, M.~{Kachelriess}, D.~V. {Semikoz}, {Explaining the Spectra of
  Cosmic Ray Groups above the Knee by Escape from the Galaxy}, ArXiv
  e-prints\href {http://arxiv.org/abs/1403.3380} {\path{arXiv:1403.3380}}.

\bibitem{JonesEllison1991}
F.~C. {Jones}, D.~C. {Ellison}, {The plasma physics of shock acceleration},
  Space Science Reviews 58 (1991) 259--346.

\bibitem{MalkovDrury2001}
M.~A. {Malkov}, L.~{O'C Drury}, {Nonlinear theory of diffusive acceleration of
  particles by shock waves}, Reports on Progress in Physics 64 (2001) 429--481.

\bibitem{LagageCesarsky:1983}
P.~O. {Lagage}, C.~J. {Cesarsky}, {The maximum energy of cosmic rays
  accelerated by supernova shocks}, Astronomy and Astrophysics 125 (1983)
  249--257.

\bibitem{Bell:2004}
A.~R. {Bell}, {Turbulent amplification of magnetic field and diffusive shock
  acceleration of cosmic rays}, MNRAS 353 (2004) 550--558.

\bibitem{DruryFalle:1986}
L.~O. {Drury}, S.~A.~E.~G. {Falle}, {On the Stability of Shocks Modified by
  Particle Acceleration}, MNRAS 223 (1986) 353.

\bibitem{BegelmanZweibel:1994}
M.~C. {Begelman}, E.~G. {Zweibel}, {Acoustic instability driven by cosmic-ray
  streaming}, Astrophys. Journal 431 (1994) 689--704.

\bibitem{DownesDrury:2012}
L.~O. {Drury}, T.~P. {Downes}, {Turbulent magnetic field amplification driven
  by cosmic ray pressure gradients}, MNRAS 427 (2012) 2308--2313.
\newblock \href {http://arxiv.org/abs/1205.6823} {\path{arXiv:1205.6823}}.

\bibitem{Bykov+2013}
A.~M. {Bykov}, A.~{Brandenburg}, M.~A. {Malkov}, S.~M. {Osipov}, {Microphysics
  of Cosmic Ray Driven Plasma Instabilities}, Space Science Reviews 178 (2013)
  201--232.
\newblock \href {http://arxiv.org/abs/1304.7081} {\path{arXiv:1304.7081}}.

\bibitem{SchureBell:2013}
K.~M. {Schure}, A.~R. {Bell}, {Magnetic field amplification by cosmic rays in
  supernova remnants}, in: G.~{Pugliese}, A.~{de Koter}, M.~{Wijburg} (Eds.),
  370 Years of Astronomy in Utrecht, Vol. 470 of Astronomical Society of the
  Pacific Conference Series, 2013, p. 209.
\newblock \href {http://arxiv.org/abs/1209.3043} {\path{arXiv:1209.3043}}.

\bibitem{Bykov+2014}
A.~M. {Bykov}, D.~C. {Ellison}, S.~M. {Osipov}, A.~E. {Vladimirov}, {Magnetic
  Field Amplification in Nonlinear Diffusive Shock Acceleration Including
  Resonant and Non-resonant Cosmic-Ray Driven Instabilities}, ApJ 789 (2014)
  137.
\newblock \href {http://arxiv.org/abs/1406.0084} {\path{arXiv:1406.0084}}.

\bibitem{DownesDrury:2014}
T.~P. {Downes}, L.~{O'C.~Drury}, {Cosmic-ray pressure driven magnetic field
  amplification: dimensional, radiative and field orientation effects}, ArXiv
  e-prints\href {http://arxiv.org/abs/1407.5664} {\path{arXiv:1407.5664}}.

\bibitem{VinkLaming:2003}
J.~{Vink}, J.~M. {Laming}, {On the Magnetic Fields and Particle Acceleration in
  Cassiopeia A}, ApJ 584 (2003) 758--769.
\newblock \href {http://arxiv.org/abs/astro-ph/0210669}
  {\path{arXiv:astro-ph/0210669}}.

\bibitem{Berezhko+2003}
E.~G. {Berezhko}, L.~T. {Ksenofontov}, H.~J. {V{\"o}lk}, {Confirmation of
  strong magnetic field amplification and nuclear cosmic ray acceleration in SN
  1006}, A\&A 412 (2003) L11--L14.
\newblock \href {http://arxiv.org/abs/astro-ph/0310862}
  {\path{arXiv:astro-ph/0310862}}.

\bibitem{Bamba+2004}
A.~{Bamba}, M.~{Ueno}, H.~{Nakajima}, K.~{Koyama}, {Thermal and Nonthermal
  X-Rays from the Large Magellanic Cloud Superbubble 30 Doradus C}, ApJ 602
  (2004) 257--263.
\newblock \href {http://arxiv.org/abs/astro-ph/0310713}
  {\path{arXiv:astro-ph/0310713}}.

\bibitem{Voelk+2005}
H.~J. {V{\"o}lk}, E.~G. {Berezhko}, L.~T. {Ksenofontov}, {Erratum: Magnetic
  field amplification in Tycho and other shell-type supernova remnants}, A\&A
  444 (2005) 893--893.

\bibitem{Parizot+2006}
E.~{Parizot}, A.~{Marcowith}, J.~{Ballet}, Y.~A. {Gallant}, {Observational
  constraints on energetic particle diffusion in young supernovae remnants:
  amplified magnetic field and maximum energy}, A\&A 453 (2006) 387--395.
\newblock \href {http://arxiv.org/abs/astro-ph/0603723}
  {\path{arXiv:astro-ph/0603723}}.

\bibitem{Uchiyama+2007}
Y.~{Uchiyama}, F.~A. {Aharonian}, T.~{Tanaka}, T.~{Takahashi}, Y.~{Maeda},
  {Extremely fast acceleration of cosmic rays in a supernova remnant}, Nature
  449 (2007) 576--578.

\bibitem{Vink:2012}
J.~{Vink}, {Supernova remnants: the X-ray perspective}, A\&A Rev. 20 (2012) 49.
\newblock \href {http://arxiv.org/abs/1112.0576} {\path{arXiv:1112.0576}}.

\bibitem{VoelkBiermann1988}
H.~J. {Voelk}, P.~L. {Biermann}, {Maximum energy of cosmic-ray particles
  accelerated by supernova remnant shocks in stellar wind cavities}, ApJ Lett.
  333 (1988) L65--L68.

\bibitem{Biermann1993}
P.~L. {Biermann}, {Cosmic rays. 1. The cosmic ray spectrum between 10' GeV and
  3 10' GeV}, A\&A 271 (1993) 649.
\newblock \href {http://arxiv.org/abs/astro-ph/9301008}
  {\path{arXiv:astro-ph/9301008}}.

\bibitem{Tatischeff2009}
V.~{Tatischeff}, {Radio emission and nonlinear diffusive shock acceleration of
  cosmic rays in the supernova SN 1993J}, A\&A 499 (2009) 191--213.
\newblock \href {http://arxiv.org/abs/0903.2944} {\path{arXiv:0903.2944}}.

\bibitem{EllisonVladimirov2008}
D.~C. {Ellison}, A.~{Vladimirov}, {Magnetic Field Amplification and Rapid Time
  Variations in SNR RX J1713.7-3946}, ApJ Lett. 673 (2008) L47--L50.
\newblock \href {http://arxiv.org/abs/0711.4389} {\path{arXiv:0711.4389}}.

\bibitem{Reeves+1970}
H.~{Reeves}, W.~{Fowler}, F.~{Hoyle}, {Galactic Cosmic Ray Origin of Li, Be and
  B in Stars}, Nature 226 (1970) 727--729.

\bibitem{MAR1971}
M.~{Meneguzzi}, J.~{Audouze}, H.~{Reeves}, {The production of the elements Li,
  Be, B by galactic cosmic rays in space and its relation with stellar
  observations.}, A\&A 15 (1971) 337--359.

\bibitem{VangioniFlam+1998}
E.~{Vangioni-Flam}, R.~{Ramaty}, K.~A. {Olive}, M.~{Casse}, {Testing the
  primary origin of Be and B in the early galaxy}, A\&A 337 (1998) 714--720.
\newblock \href {http://arxiv.org/abs/astro-ph/9806084}
  {\path{arXiv:astro-ph/9806084}}.

\bibitem{Gilmore+1992}
G.~{Gilmore}, B.~{Gustafsson}, B.~{Edvardsson}, P.~E. {Nissen}, {Is beryllium
  in metal-poor stars of galactic or cosmological origin?}, Nature 357 (1992)
  379--384.

\bibitem{Ryan+1992}
S.~G. {Ryan}, J.~E. {Norris}, M.~S. {Bessell}, C.~{Deliyannis}, {Evolution of
  beryllium abundances in the galactic halo}, ApJ 388 (1992) 184--189.

\bibitem{BoesgaardKing1993}
A.~M. {Boesgaard}, J.~R. {King}, {Galactic evolution of Beryllium},
  Astronomical Journal 106 (1993) 2309--2323.

\bibitem{Duncan+1992}
D.~K. {Duncan}, D.~L. {Lambert}, M.~{Lemke}, {The abundance of boron in three
  halo stars}, ApJ 401 (1992) 584--595.

\bibitem{Ryan+1996}
S.~G. {Ryan}, J.~E. {Norris}, T.~C. {Beers}, {Extremely Metal-poor Stars. II.
  Elemental Abundances and the Early Chemical Enrichment of the Galaxy}, ApJ
  471 (1996) 254.

\bibitem{Walker+1993}
T.~P. {Walker}, G.~{Steigman}, D.~N. {Schramm}, K.~A. {Olive}, B.~{Fields},
  {The boron-to-beryllium ratio in halo stars - A signature of cosmic-ray
  nucleosynthesis in the early Galaxy}, ApJ 413 (1993) 562--570.

\bibitem{Ramaty+1997}
R.~{Ramaty}, B.~{Kozlovsky}, R.~E. {Lingenfelter}, H.~{Reeves}, {Light Elements
  and Cosmic Rays in the Early Galaxy}, ApJ 488 (1997) 730--748.
\newblock \href {http://arxiv.org/abs/astro-ph/9610255}
  {\path{arXiv:astro-ph/9610255}}.

\bibitem{ParizotDrury1999a}
E.~{Parizot}, L.~{Drury}, {Spallative nucleosynthesis in supernova remnants. I.
  Analytical estimates}, A\&A 346 (1999) 329--339.
\newblock \href {http://arxiv.org/abs/astro-ph/9811471}
  {\path{arXiv:astro-ph/9811471}}.

\bibitem{ParizotDrury1999b}
E.~{Parizot}, L.~{Drury}, {Spallative nucleosynthesis in supernova remnants.
  II. Time-dependent numerical results}, A\&A 346 (1999) 686--698.
\newblock \href {http://arxiv.org/abs/astro-ph/9903358}
  {\path{arXiv:astro-ph/9903358}}.

\bibitem{Castor+1975}
J.~{Castor}, R.~{McCray}, R.~{Weaver}, {Interstellar bubbles}, ApJ Lett. 200
  (1975) L107--L110.

\bibitem{Weaver+1977}
R.~{Weaver}, R.~{McCray}, J.~{Castor}, P.~{Shapiro}, R.~{Moore}, {Interstellar
  bubbles. II - Structure and evolution}, ApJ 218 (1977) 377--395.

\bibitem{MacLowMcCray1988}
M.-M. {Mac Low}, R.~{McCray}, {Superbubbles in disk galaxies}, ApJ 324 (1988)
  776--785.

\bibitem{MacLow+1989}
M.-M. {Mac Low}, R.~{McCray}, M.~L. {Norman}, {Superbubble blowout dynamics},
  ApJ 337 (1989) 141--154.

\bibitem{ChuMacLow1990}
Y.-H. {Chu}, M.-M. {Mac Low}, {X-rays from superbubbles in the Large Magellanic
  Cloud}, ApJ 365 (1990) 510--521.

\bibitem{Ferriere+1991}
K.~M. {Ferriere}, M.-M. {Mac Low}, E.~G. {Zweibel}, {Expansion of a superbubble
  in a uniform magnetic field}, ApJ 375 (1991) 239--253.

\bibitem{BaumBreit2013}
V.~{Baumgartner}, D.~{Breitschwerdt}, {Superbubble evolution in disk galaxies.
  I. Study of blow-out by analytical models}, A\&A 557 (2013) A140.
\newblock \href {http://arxiv.org/abs/1402.0194} {\path{arXiv:1402.0194}}.

\bibitem{Higdon+1998}
J.~C. {Higdon}, R.~E. {Lingenfelter}, R.~{Ramaty}, {Cosmic-Ray Acceleration
  from Supernova Ejecta in Superbubbles}, ApJL 509 (1998) L33--L36.

\bibitem{HigdonLingenfelter2005}
J.~C. {Higdon}, R.~E. {Lingenfelter}, {OB Associations, Supernova-generated
  Superbubbles, and the Source of Cosmic Rays}, ApJ 628 (2005) 738--749.

\bibitem{BykovFleishman1992}
A.~M. {Bykov}, G.~D. {Fleishman}, {On non-thermal particle generation in
  superbubbles}, MNRAS 255 (1992) 269--275.

\bibitem{Bykov+1995}
A.~M. {Bykov}, V.~S. {Ptuskin}, I.~N. {Toptygin}, {Spectrum of Ultra-High
  Energy Cosmic Rays Acceleration in Superbubbles}, International Cosmic Ray
  Conference 3 (1995) 337.

\bibitem{BykovToptygin2001}
A.~M. {Bykov}, I.~N. {Toptygin}, {A Model of Particle Acceleration to High
  Energies by Multiple Supernova Explosions in OB Associations}, Astronomy
  Letters 27 (2001) 625--633.

\bibitem{Bykov2001}
A.~M. {Bykov}, {Particle Acceleration and Nonthermal Phenomena in
  Superbubbles}, Space Science Reviews 99 (2001) 317--326.

\bibitem{Parizot+2004}
E.~{Parizot}, A.~{Marcowith}, E.~{van der Swaluw}, A.~M. {Bykov},
  V.~{Tatischeff}, {Superbubbles and energetic particles in the Galaxy. I.
  Collective effects of particle acceleration}, A\&A 424 (2004) 747--760.
\newblock \href {http://arxiv.org/abs/astro-ph/0405531}
  {\path{arXiv:astro-ph/0405531}}.

\bibitem{ButtBykov2008}
Y.~M. {Butt}, A.~M. {Bykov}, {A Cosmic-Ray Resolution to the Superbubble Energy
  Crisis}, ApJ Lett. 677 (2008) L21--L22.
\newblock \href {http://arxiv.org/abs/0802.3805} {\path{arXiv:0802.3805}}.

\bibitem{ParizotDrury1999}
E.~{Parizot}, L.~{Drury}, {Superbubbles as the source of (6) Li, Be and B in
  the early Galaxy}, A\&A 349 (1999) 673--684.
\newblock \href {http://arxiv.org/abs/astro-ph/9906298}
  {\path{arXiv:astro-ph/9906298}}.

\bibitem{ParizotDrury2000a}
E.~{Parizot}, L.~{Drury}, {The Superbubble Model for LiBeB Production and
  Galactic Evolution}, in: L.~{da Silva}, R.~{de Medeiros}, M.~{Spite} (Eds.),
  The Light Elements and their Evolution, Vol. 198 of IAU Symposium, 2000,
  p.~35.
\newblock \href {http://arxiv.org/abs/astro-ph/0002208}
  {\path{arXiv:astro-ph/0002208}}.

\bibitem{ParizotDrury2000b}
E.~{Parizot}, L.~{Drury}, {Bimodal production of Be and B in the early Galaxy},
  A\&A 356 (2000) L66--L70.
\newblock \href {http://arxiv.org/abs/astro-ph/0003026}
  {\path{arXiv:astro-ph/0003026}}.

\bibitem{Parizot2001}
E.~{Parizot}, {Galactic Cosmic Rays and the Light Elements}, Space Science
  Reviews 99 (2001) 61--71.

\bibitem{Maehl+1975}
R.~{Maehl}, F.~A. {Hagen}, A.~J. {Fisher}, J.~F. {Ormes}, M.~{Simon},
  {Astrophysical implications of the isotopic composition of cosmic rays},
  International Cosmic Ray Conference 1 (1975) 367--372.

\bibitem{GarciaMunoz+1979}
M.~{Garcia-Munoz}, J.~A. {Simpson}, J.~P. {Wefel}, {The isotopes of neon in the
  galactic cosmic rays}, ApJ Lett. 232 (1979) L95--L99.

\bibitem{Duvernois+1996}
M.~A. {Duvernois}, M.~{Garcia-Munoz}, K.~R. {Pyle}, J.~A. {Simpson}, M.~R.
  {Thayer}, {The Isotopic Composition of Galactic Cosmic-Ray Elements from
  Carbon to Silicon: The Combined Release and Radiation Effects Satellite
  Investigation}, ApJ 466 (1996) 457.

\bibitem{Webber+1997}
W.~R. {Webber}, A.~{Lukasiak}, F.~B. {McDonald}, {Voyager Measurements of the
  Mass Composition of Cosmic-Ray Ne, Mg, Si and S Nuclei}, ApJ 476 (1997)
  766--770.

\bibitem{Binns+2006a}
W.~R. {Binns}, M.~E. {Wiedenbeck}, M.~{Arnould}, {et al.}, {Wolf-Rayet stars,
  OB associations, and the origin of galactic cosmic rays}, New Astronomy
  Reviews 50 (2006) 516--520.

\bibitem{Binns+2006b}
W.~R. {Binns}, M.~E. {Wiedenbeck}, M.~{Arnould}, {et al.}, {Superbubbles,
  Wolf-Rayet stars, and the origin of galactic cosmic rays}, Journal of Physics
  Conference Series 47 (2006) 68--77.

\bibitem{HigdonLingenfelter2003}
J.~C. {Higdon}, R.~E. {Lingenfelter}, {The Superbubble Origin of $^{22}$Ne in
  Cosmic Rays}, Astrophysical Journal 590 (2003) 822--832.

\bibitem{HigdonLingenfelter2006}
J.~C. {Higdon}, R.~E. {Lingenfelter}, {The superbubble origin for galactic
  cosmic rays}, Advances in Space Research 37 (2006) 1913--1917.

\bibitem{LingenfelterHigdon2007}
R.~E. {Lingenfelter}, J.~C. {Higdon}, {The Composition of Cosmic Rays and the
  Mixing of the Interstellar Medium}, Space Science Reviews 130 (2007)
  465--473.

\bibitem{Meyer+1997}
J.-P. {Meyer}, L.~O. {Drury}, D.~C. {Ellison}, {Galactic Cosmic Rays from
  Supernova Remnants. I. A Cosmic-Ray Composition Controlled by Volatility and
  Mass-to-Charge Ratio}, ApJ 487 (1997) 182--196.
\newblock \href {http://arxiv.org/abs/astro-ph/9704267}
  {\path{arXiv:astro-ph/9704267}}.

\bibitem{Ramaty+1979a}
R.~{Ramaty}, R.~E. {Lingenfelter}, {Gamma-ray line astronomy}, Nature 278
  (1979) 127--132.

\bibitem{Ramaty+1979b}
R.~{Ramaty}, B.~{Kozlovsky}, R.~E. {Lingenfelter}, {Nuclear gamma-rays from
  energetic particle interactions}, ApJS 40 (1979) 487--526.

\bibitem{Tibolla+2011a}
O.~{Tibolla}, K.~{Mannheim}, A.~{Summa}, A.~{Paravac}, J.~{Greiner},
  G.~{Kanbach}, {Nuclear lines revealing the injection of cosmic rays in
  supernova remnants}, ArXiv e-prints\href {http://arxiv.org/abs/1106.1023}
  {\path{arXiv:1106.1023}}.

\bibitem{Tibolla+2011b}
O.~{Tibolla}, K.~{Mannheim}, A.~{Paravac}, J.~{Greiner}, G.~{Kanbach}, {GRIPS
  and its strong connections to the GeV and TeV bands}, Nuovo Cimento C
  Geophysics Space Physics C 34 (2011) 41--47.
\newblock \href {http://arxiv.org/abs/1106.1012} {\path{arXiv:1106.1012}}.

\bibitem{Casse+1995a}
M.~{Cass{\'e}}, R.~{Lehoucq}, E.~{Vangioni-Flam}, {Production and evolution of
  light elements in active star-forming regions}, Nature 373 (1995) 318--319.

\bibitem{Roy+2013}
A.~{Roy}, B.~B. {Nath}, P.~{Sharma}, Y.~{Shchekinov}, {Superbubble breakout and
  galactic winds from disc galaxies}, MNRAS 434 (2013) 3572--3581.
\newblock \href {http://arxiv.org/abs/1303.2664} {\path{arXiv:1303.2664}}.

\bibitem{BertainaParizot:2014}
M.~{Bertaina}, E.~{Parizot}, {The JEM-EUSO mission: a space observatory to
  study the origin of Ultra-High Energy Cosmic Rays}, in: O.~{Tibolla},
  L.~{Drury} (Eds.), Cosmic Ray Origin beyond the Standard Model, this volume,
  2014.

\bibitem{Lebrun+2003}
F.~{Lebrun}, A.~{Bazzano}, V.~{Borrel}, {et al.}, {The Compton Cube}, Nuclear
  Instruments and Methods in Physics Research A 504 (2003) 38--43.

\end{thebibliography}







\end{document}